# A Compressed Sensing Framework of Frequency-Sparse Signals through Chaotic Systems

Z. Liu, S.Y. Chen and F. Xi,

Department of Electronic Engineering

Nanjing University of Science and Technology

Nanjing, Jiangsu 210094

People's Republic of China

eezliu@mail.njust.edu.cn

## ABSTRACT

This paper proposes a compressed sensing (CS) framework for the acquisition and reconstruction of frequency-sparse signals with chaotic dynamical systems. The sparse signal is acting as an excitation term of a discrete-time chaotic system and the compressed measurement is obtained by downsampling the system output. The reconstruction is realized through the estimation of the excitation coefficients with principle of impulsive chaos synchronization. The $l_1$-norm regularized nonlinear least squares is used to find the estimation. The proposed framework is easily implementable and creates secure measurements. The Henon map is used as an example to illustrate the principle and the performance.

## I. INTRODUCTION

Compressed sensing (CS), introduced by Candes and Tao [Candes & Tao, 2006] and Donoho [Donoho, 2006], is an emerging framework for representing sparse signals, *i.e.*, the signals that are well approximated by a short linear combination of vectors from a basis or a dictionary. With the theory, the sparse signals can be captured from relatively few measurements, typically well below that required by the Shannon/Nyquist sampling theorem. The essence in CS theory is to create the measurements through a random linear projection without *a priori* knowledge of the signal model. The sparsity assumption is the necessity for reconstructing the signal. The reconstruction is often done through $l_1$ optimization-based algorithms [Candes *et al*., 2008; Daubechies *et al*., 2010]. For incoherent measurements, high reconstruction probability will be achieved.

As an alternative paradigm, random FIR filters [Tropp *et al*., 2006; Tropp, 2006] have been proposed for compressive signal acquisition (abbr. as RanCS in this paper). As shown in Fig. 1, the compressive measurement is conducted by convolving the signal with a random-tap FIR filter and then downsampling the filter output. The filter taps are randomly chosen from some kind of distributions. It is shown that the random filters are sufficiently generic to summarize many types of sparse signals. The random filtering and downsampling operation functions like the linear projection in CS and the downsampling output can be represented as a linear measurement on the input signal. Then the reconstruction is implemented as in usual CS reconstruction.

Until now, the CS theory has been mainly developed with linear measurements. While simple and tractable in mathematics, there are some practical difficulties in the CS implementation. For high-dimensional data, the measurement system often results in large computational load for sensing outputs. In remote applications, the large measurement system data have to be transmitted to the receiver for the reconstruction.

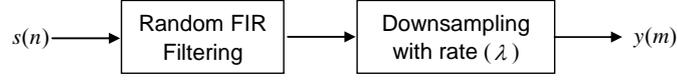

Fig.1 Random Filter Structure for CS

This paper proposes a chaos-based CS structure (termed as ChaCS), which overcomes the limitations of current CS theory to some extent. As shown in Fig.2(a), ChaCS performs the measurement by firstly passing the signal through a discrete-time chaotic system (Master Chaotic System) and then downsampling the system output. The chaotic systems play a role similar to random filters in Fig.1. In general, simple chaotic systems will behave like a large size of random filters and therefore, the measurement is implementable with simple hardware. The reconstruction is done through impulsive synchronization principle in conjunction with parameter estimation techniques, as shown in Fig.2(b). Upon synchronizing, the excitation parameters are estimated and therefore, the sparse signal is reconstructed. The impulsive synchronization theory is fundamental of the proposed structure. In chaotic dynamics, this is a well-developed area. See [Yang & Chua, 1997; Itoh *et al*, 2001; Liu, 2001] and references therein for details. For our application, the downsampling outputs are acting as the synchronization impulses and are employed to control the parameter estimation.

In the proposed ChaCS, the sparse signals act as the excitations to the chaotic systems. For frequency-sparse signals, the excitations will last during the measurement intervals. However, for time-sparse signals, the excitations will have effect only on the initial states of the chaotic systems. If the systems have short transient process and the downsampling rates are high enough, the downsampling outputs will have no information of the excitation signals. Therefore, the ChaCS is mostly suitable for frequency-sparse signals, which are also ones that the current paper focuses on. Similar to RanCS, the ChaCS is also suited for real-time applications.

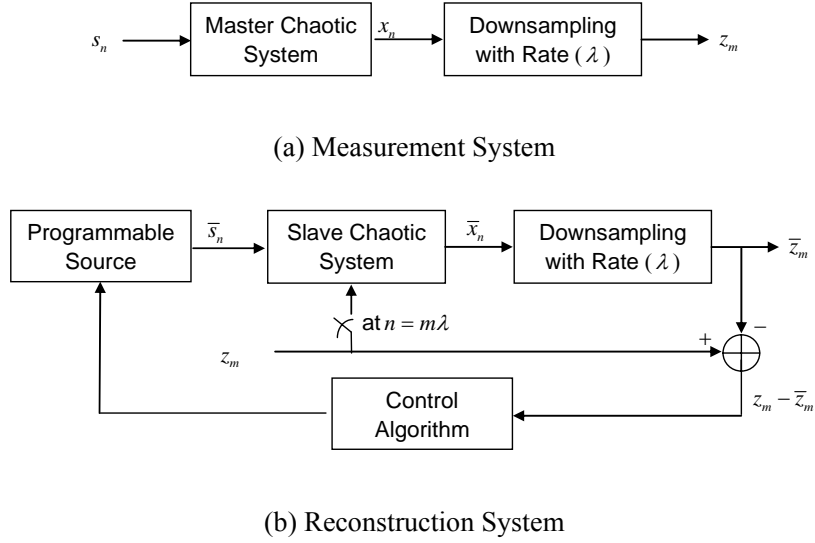

(a) Measurement System

(b) Reconstruction System

Fig.2 Chaos-Based Structure for CS

As for the applications of chaos to CS theory, some attempts [Nguyen *et al*, 2008; Yu *et al*, 2010] have been made. All of these works are based on conventional CS theory and focus on the generation of the measurement matrix using chaotic sequences. A work related to that in the paper is the CS with nonlinear observations [Blumensath, 2010]. The author introduces the concept of nonlinear measurements into CS theory and derives the sufficient reconstruction conditions from the linearization of the maps. However, the proposed ChaCS is derived from different mechanisms. The reconstruction is guaranteed by the impulsive chaos synchronization.

The current work involves CS theory, chaos theory, chaotic synchronization, and parameter estimation and so on. It is not intended to detail all materials in this paper. Some key references will be referred to for related materials. We will take the Henon map [Gleick, 1987] as an example to illustrate the chaos-based CS principle and its performance.

## II. IMPULSIVE SYNCHRONIZATION OF CHAOTIC SYSTEMS

Impulsive synchronization theory shows that two chaotic systems can be synchronized through coupling of impulse samples from some state variables of the systems. For the unidirectional coupling, the impulse samples from one of the systems (master chaotic system) are acting on another system (slave chaotic system) at the sample instants. Let us take the discrete-time Henon map as an example to illustrate the principle.

The master system is described by

$$\begin{cases} x_{n+1} = -ax_n^2 + y_n + 1 \\ y_{n+1} = bx_n \end{cases} \quad (1)$$

where $a$ and $b$ are system parameters. With proper setting on the parameters, the system will be running in chaotic states. One example is shown in Fig.3 for $(a,b) = (1.4, 0.3)$ and $(x_0, y_0) = (0.25, 0.25)$.

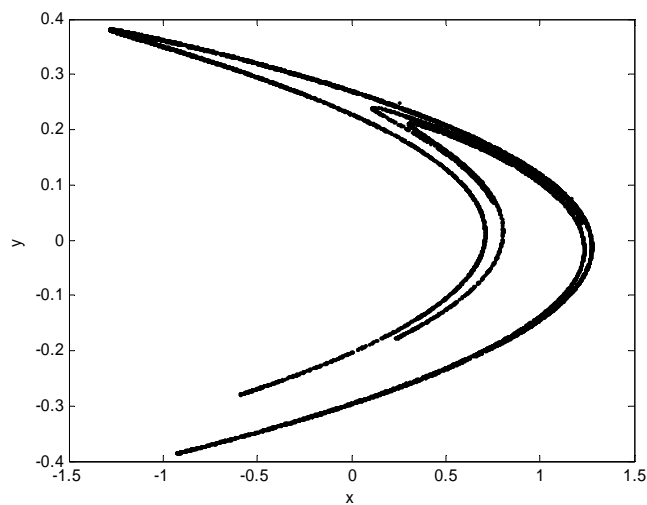

Fig.3 The Henon map attractor

For equal-spacing impulse sampling, the slave system is given by

$$\begin{cases} \bar{x}_{n+1} = x_{n+1} & n = \lambda m \\ \bar{x}_{n+1} = -a\bar{x}_n^2 + \bar{y}_n + 1, & n \neq \lambda m \\ \bar{y}_{n+1} = b\bar{x}_n \end{cases} \quad (2)$$

where $\lambda$ is an integer and $m = 0,1,2,\cdots$. It is seen from (2) that the driving state $x_n$ is acting on the slave system only at time instants $n = \lambda m$, while the slave system is running freely at other time instants. With variational error system, it can be shown that (1) and (2) will be synchronized to each other for suitable $\lambda$, i.e., $(\bar{x}_n, \bar{y}_n) \to (x_n, y_n)$ as $n \to \infty$. For $(a,b) = (1.4, 0.3)$, (2) can be synchronized to (1) with $\lambda$ up to 4. A synchronization process is shown in Fig.4. The synchronization is called impulsive synchronization in chaos synchronization theory.

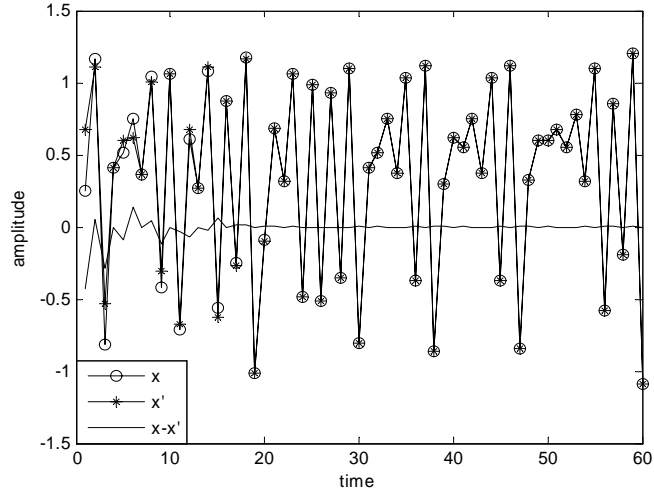

Fig.4 The impulsive synchronization process

Different from the conventional coupling synchronization, the driving information from the master system is coupled to the slave system only at selective time instants. It is the driving way that makes the compressed measurements possible as shown in next section.

## III. ChaCS THEORY

### 3.1 ChaCS Measurement

Consider the Henon map excited by the compressive signal $s_n$

$$\begin{cases} x_{n+1} = -ax_n^2 + y_n + 1 \\ y_{n+1} = bx_n + s_n \end{cases} \quad (3)$$

Suppose that $s_n$ is expressed as $s_n = \sum_{k=1}^{N} \alpha_k \phi_{n,k}$, where $\{\phi_{n,k}\}_{k=1}^{N}$ are real Fourier basis signals and $\{\alpha_k\}_{k=1}^{N}$ are expansion coefficients. For finite-length $s_n$, there are only $K$ nonzero coefficients with $K \ll N$. For infinite-length signal, the $K$ is small in comparison with the corresponding signal bandwidth [Tropp et al, 2010].

Let $x_n$ be observable. The measurements $z_m$ of $s_n$ are defined as the downsampling of $x_n$,

$$z_m = D_\downarrow(x_n) \quad (4)$$

where $n = \lambda m$ and $\lambda$ is here referred to as the downsampling rate. For signal of length $N$, the $M = \lfloor N/\lambda \rfloor$ measurements will be taken. The measurement principle is illustrated in Fig.2(a).

The essential idea in CS theory is to "randomize" the sensing signal through a random projection. The randomization in the ChaCS is achieved through chaotic process. The amplitude of the signal $s_n$ is assumed to be small so that it does not destroy the chaotic behaviors of the system (3). Otherwise, (3) is acting as a linear or nonlinear filter and it is not enough to capture the compressive signal using the downsampling outputs. For chaotic systems, the system output is random-like and is similar to that produced by random filters. As a by-product, the measurement system creates secure measurements because of chaotization of the input signals.

**3.2 ChaCS Reconstruction**

To perform the reconstruction of $s_n$ (or estimation of $\{\alpha_k\}_{k=1}^{N}$) from $z_m$, let us consider slave system as

$$\begin{cases} \overline{x}_{n+1} = x_n & n = \lambda m \\ \overline{x}_{n+1} = -a\overline{x}_n^2 + \overline{y}_n + 1, & n \neq \lambda m \\ \overline{y}_{n+1} = b\overline{x}_n + \overline{s}_n \end{cases} \quad (5)$$

where $\overline{s}_n = \sum_{k=1}^{N} \overline{\alpha}_k \phi_{n,k}$ and $\{\overline{\alpha}_k\}_{k=1}^{N}$ is unknown. Similar to (4), we can generate

$$\overline{z}_m = D_\downarrow(\overline{x}_n) \quad (6)$$

If $\{\overline{\alpha}_k\}_{k=1}^{N} = \{\alpha_k\}_{k=1}^{N}$, (5) will be synchronized to (3) with synchronizable $\lambda$, *i.e.*, $(\overline{x}_n, \overline{y}_n) \to (x_n, y_n)$ as $n \to \infty$.

In CS application, the initial conditions for the system (3) can be passed to the slave system (5). Then $\{\overline{\alpha}_k\}_{k=1}^{N}$ can be adjusted to $\{\alpha_k\}_{k=1}^{N}$ upon the synchronization, *i.e.*, $(\overline{x}_n, \overline{y}_n) \to (x_n, y_n)$ during the observation time interval. With $z_m$ and $\overline{z}_m$ available, the impulsive synchronization-based ChaCS reconstruction can be constructed as shown in Fig.2(b). The error outputs between $z_m$ and $\overline{z}_m$ are used to formulate the updating algorithm of $\{\overline{\alpha}_k\}_{k=1}^{N}$, which adjusts the sensing signal in the slave system. Upon synchronizing, $\overline{z}_m \to z_m$ and $\{\overline{\alpha}_k\}_{k=1}^{N} \to \{\alpha_k\}_{k=1}^{N}$.

With $z_m$ from the master system and $\overline{z}_m$ from the slave system, the problem is to design an updating rule for estimating $\{\overline{\alpha}_k\}_{k=1}^{N}$ such that $\overline{z}_m = z_m$. The unknown $\{\bar{\alpha}_k\}_{k=1}^N$ is implicitly included in $\bar{z}_m$. This can be done by nonlinear least squares formulation as

$$E = \sum_{m=1}^{M} |z_m - \bar{z}_m|^2 \qquad (7)$$

Some well-developed algorithms [Madsen *et al*, 2004] can be used to find the estimate. However, when $M$, the number of measurements, is not large enough compared to $N$, the unknown number, the simple least-squares minimization (7) may leads to over-fit. Considering the sparsity in estimation parameters, as in conventional CS reconstruction, the regularization techniques can be used to enhance the sparsity. One of the formulations is to do $l_1$-regularized least-squares,

$$\min_{\bar{\alpha}} \sum_{m=1}^{M} |z_m - \bar{z}_m|^2 + \mu \|\bar{\alpha}\|_1 \qquad (8)$$

where $\bar{\alpha} = [\bar{\alpha}_1, \bar{\alpha}_2, \cdots, \bar{\alpha}_N]^T$ and $\mu > 0$ is the regularization parameter. Solving the (8) has attracted attention recently and some algorithms are suggested [Tseng & Yun, 2009; Schmidt *et al*., 2007; Schmidt *et al*., 2009]. Being highly efficient, the iteratively re-weighted least squares (IRLS) algorithm [Daubechies *et al*, 2010] is generalized in this paper as iteratively re-weighted nonlinear least squares (IRNLS) one to solve (8). The generalization is done by replacing the linear least squares estimation with nonlinear least squares one. An algorithmic framework is given in appendix.

It must be noted that both formulations (7) and (8) are nonlinear and nonconvex because of nonlinear properties in chaotic systems and all of the algorithms are possible to produce the local minima. Our simulations in next section demonstrate that the local minima may derive the reasonable reconstruction.

In conventional CS measurement or sampling using structure in Fig.1(a), the number of measurements or downsampling rate depends on several components, for example, sparsity level, signal length, coherence between the measurement vector and the sparsity basis [Tropp *et al.*, 2006]. If the random measurement does not satisfy the restricted isometry property conditions [Candes & Tao, 2006], the reconstruction cannot be ensured with 100% success for every sparse signal. For the proposed ChaCS, the number of measurements will depend only on width of synchronizable impulse samples for the parameter identifiable chaotic systems[1]. In theory, for the impulsively synchronized chaotic system, we could ensure 100% successful reconstruction if the parameters were identifiable and a global minimum of (7) or (8) could be found. However, because of the high nonlinearity in formulation (8) and the lack of the global solvers for the high sparsity level, the measurement number will also depend on the sparsity level as usual CS.

## IV. SIMULATION EXPERIMENTS

For the proposed ChaCS, we have conducted the extensive simulations to demonstrate its performance. Some samples with finite-length signals are shown here, which show that the chaotic systems are powerful for compressed sampling and the local minima produce the reasonable reconstruction of the sensing signals.

The simulation scenarios are set up as follows: The frequency-sparse signals are generated with real Fourier bases. The sparse positions are uniformly distributed over digital frequency range [0, 0.5]. The sparse coefficients are from two types of distributions: *i.i.d.* Gaussian with zero mean and unity variance and *i.i.d.* Bernoulli with entries $\pm 1$. Different sparsity levels and downsampling rates are considered. The Henon parameters are set as $(a,b)=(1.4, 0.3)$. The sensing signals are scaled such that the excited Henon maps are in chaotic states.

---

[1] This section is to describe the general principle of the ChaCS. It is not clear if the outputs of the master chaotic system carry enough information to perform the reconstruction. This is an identifiability problem in nonlinear systems. Some methods [Dedieu & Ogorzalek, 1997; Floriane *et al.*, 2008] may be used to check the identifiability. However, it is needed to do further research for the identifiability of chaotic systems based on impulsive chaos synchronization. In our discussion, we assume that the parameters are identifiable.

The IRNLS algorithm in appendix is used to compute the sparse coefficients. The initial setting of each coefficient for IRNLS is randomly chosen over interval [-1, 1]. Other parameters $\mu$ and $\varepsilon$ are set to be $10^{-6}$ and $10^{-14}$, respectively. The IRNLS is deemed convergent if the relative error (stopping criterion) between two consecutive iterations is less than $10^{-5}$. The solution by the IRNLS with one set of initial settings is called one realization. The median relative error is to measure reconstruction performance of the sparse signals,

$$Err = \sum_{n=1}^{N}(x_n - \bar{x}_n)^2 \bigg/ \sum_{n=1}^{N} x_n^2$$

Figures 5 and 6 show one realization for the case of the Gaussian coefficients with $\lambda = 2$ and $N = 128$. Two sparsity levels ($K = 15$ and $K = 22$) are simulated. Sparsity positions and amplitudes are shown in Fig.5 (a) and Fig.6 (a), respectively. It is seen from Fig.5 that for $K = 15$ the reconstructed signal matches the original one well and sparse positions/coefficients are estimated correctly. The estimated relative error is $3.89 \times 10^{-6}$. In this case, the global minimum is found. While for $K = 22$, only the local minimum is found and the sparse positions/coefficients are not well estimated, as seen from Fig.6 (a). However, the reconstructed signal still is a reasonable approximation to the original one (Fig.6(b)). From the point of view of waveform reconstruction, the sparse signal is well reconstructed, although the local optimal coefficients are found.

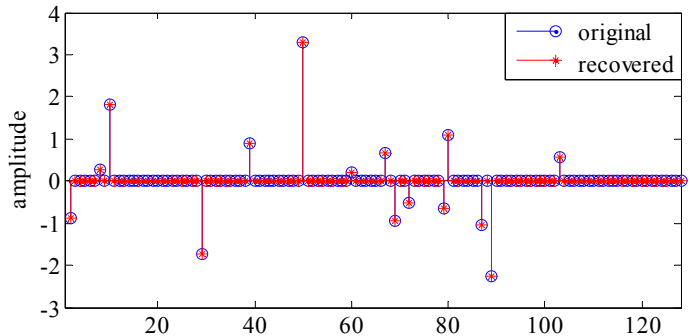

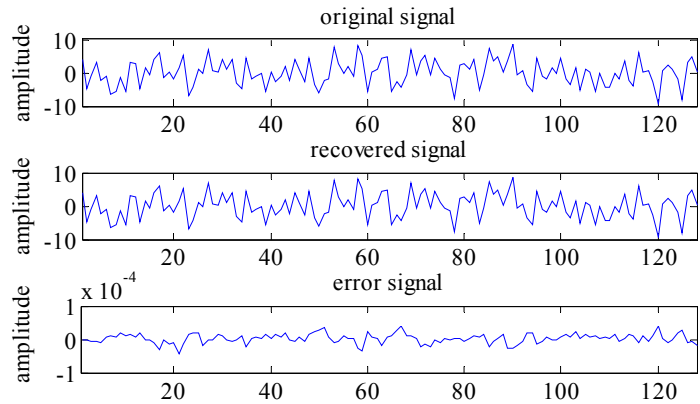

(a) Sparsity and its estimation

(b) Original waveform and its reconstruction

Fig.5 Sparse signal and its reconstruction for $K=15$ with Gaussian coefficients

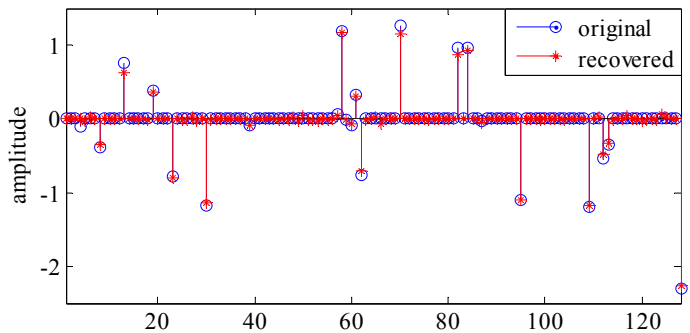

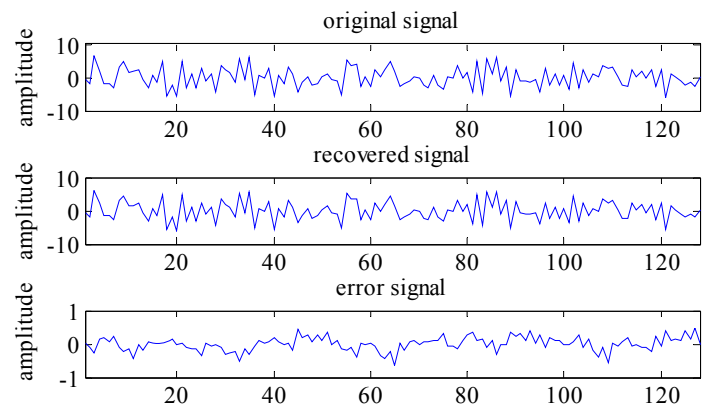

(a) Sparsity and its estimation

(b) Original waveform and its reconstruction

Fig.6 Sparse signal and its reconstruction $K=22$ with Gaussian coefficients
12

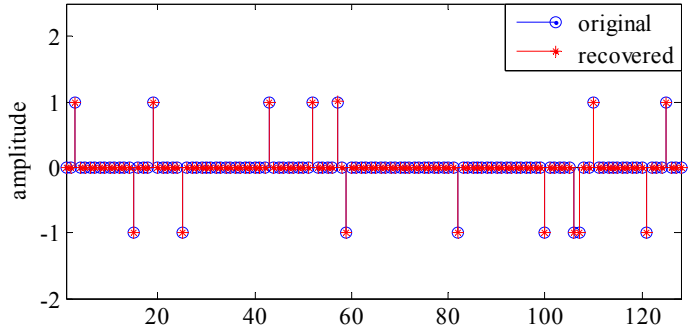

(a) Sparsity and its estimation

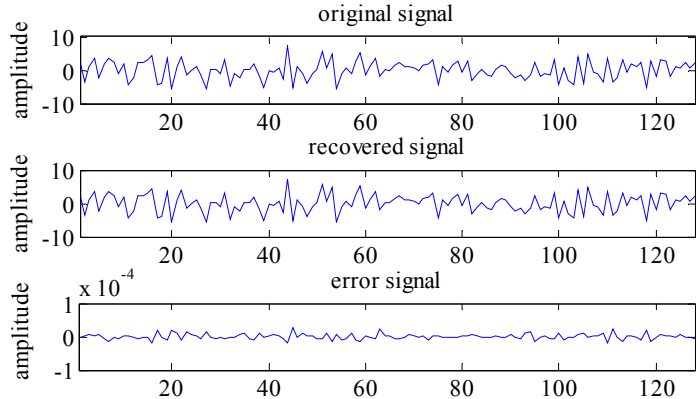

(b) Original waveform and its reconstruction

Fig.7 Sparse signal and its reconstruction for $K=15$ with Bernoulli coefficients

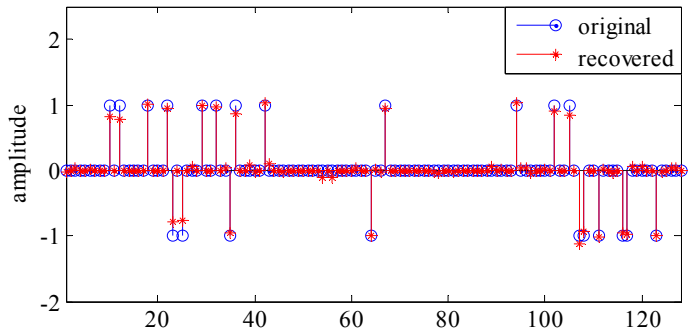

(a) Sparsity and its estimation

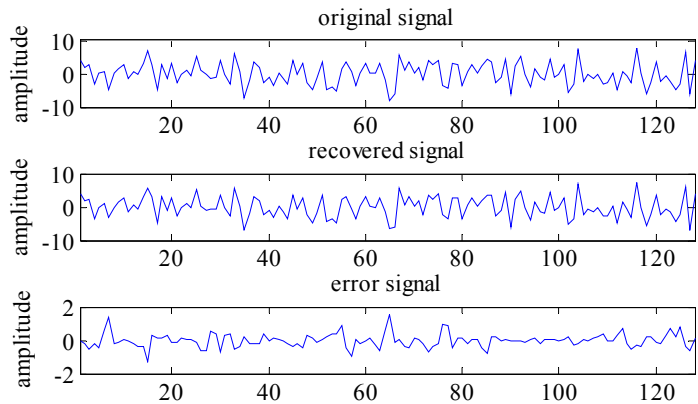

(b) Original waveform and its reconstruction

Fig.8 Sparse signal and its reconstruction $K=22$ with Bernoulli coefficients

Figures 7 and 8 show the results similar to those in Fig.5 and Fig.6 for the Bernoulli coefficients. For the two sparsity levels, the same conclusions as in Fig.5 and Fig.6 can be drawn.

The results in Figs.5~8 also declare that the sparsity has effect on the global minimum. To further evaluate the effect of sparsity on the reconstruction of sparse signals, we conduct different experiments with different initial settings for coefficients. The averaged results over 100 realizations are shown in Fig.9 for two coefficient distributions. As seen, the estimation performance decreases as the sparsity increases. The downsampling rates (width of impulse sampling) have the same effects as the sparsity.

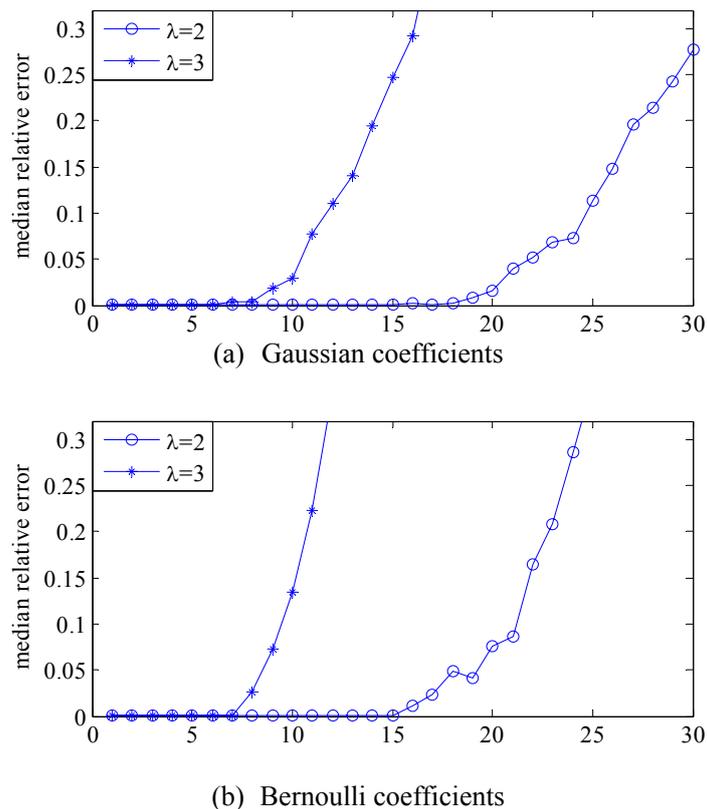

(b) Bernoulli coefficients
Fig.9 Median relative errors vs. sparsity of the proposed ChaCS structure
with different downsampling rates

As for two coefficient distributions, the reconstruction performance of the Gaussian distribution is superior to that of Bernoulli one. The performance discrepancy comes from the IRNLS algorithm itself. Similar to the reweighted $l_1$ algorithm in [Candes *et al*., 2008], the large coefficients in Gaussian distribution are easily identified and significantly downweighted early in the reweighting process. However, for the Bernoulli coefficients, there is no such superiority.

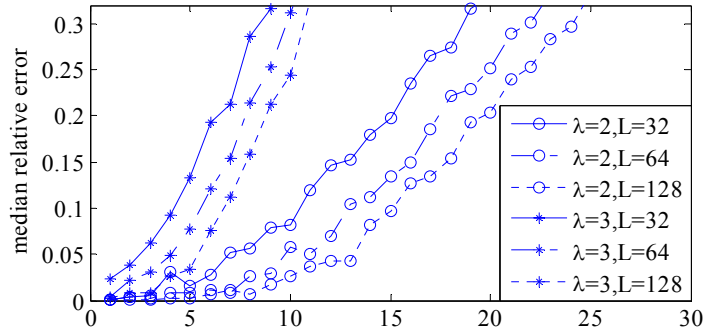

(a) Gaussian coefficients

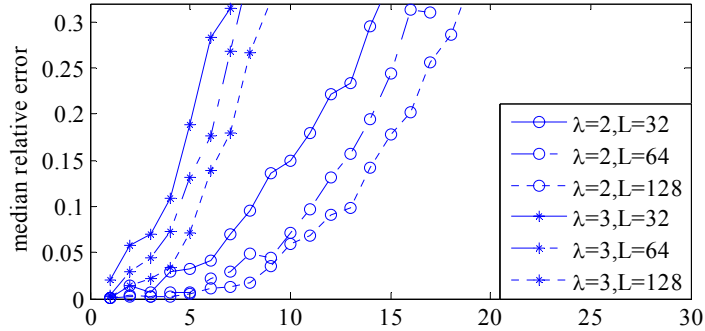

(b) Bernoulli coefficients

Fig.10 Median relative errors vs. sparsity of the RanCS structre
with different filter lengths (L's) and downsampling rates

To make comparison with the RanCS structure in Fig.1, we have done simulation experiments as in Fig. 9 with different filter lengths (L's). The filter coefficients are randomly chosen with $N(0,1)$ Gaussian distribution. The IRLS algorithm [Daubechies *et al*, 2010] is used to perform the reconstruction. The results are shown in Fig.10. It is seen that the proposed ChaCS structure with the specific chaotic map shows the performance superior to that by the RanCS structure.

## V. CONCLUSION

This paper has developed a new framework for compressed sensing using chaos synchronization theory. The compressed measurements are obtained by downsampling outputs of chaotic system excited by sensing signal. The reconstruction is based on chaos synchronization principle with parameter estimation techniques. Simulation results show the performance superiority of the proposed framework realized by Henon map.

In comparison with conventional CS, the measurement system of the proposed ChaCS is simple and easily implementable, but the reconstruction system is complex. Varieties of chaotic systems provide much freedom for the measurement implementation.

The idea of using chaotic systems to CS is appealing and needs to be further studied. The core of conventional CS theories is directly related to Gelfand widths and development of $l_1$-optimization. However, the new structure is mainly built upon the chaos theory, which may provide us new sight on development of CS theory.

**APPENDIX**

In this appendix, we generalize the IRLS algorithm [Daubechies *et al*, 2010] to solving (8). The fundamental idea is to replace the $l_1$-regularized term in (8) by a weighted $l_2$ norm with the objective function as,

$$\min_{\bar{\alpha}} \sum_{m=1}^{M} |z_m - \bar{z}_m|^2 + \mu \|W^{1/2} \bar{\alpha}\|_2 \tag{A1}$$

where the weights are computed from the previous iterate $\bar{\alpha}^{j-1}$ by $w_k^j = ((\bar{\alpha}_k^j)^2 + \varepsilon)^{-1/2}$ $(k = 1, 2, \cdots, N)$. $\varepsilon$ is a small positive number in case of some $\bar{\alpha}_k^{j-1} = 0$. Then the term $\|W^{1/2}\bar{\alpha}\|_2$ in (A1) approximates $\|\bar{\alpha}\|_1$ in a first-order. The algorithmic framework is described as follows:

**Algorithm** IRNLS

**Objective**: Minimize (8) with respect to $\bar{\alpha}$

**Input**: $\mu, \varepsilon$, $err$ (stopping criterion).

**Initialization**: Set $\bar{\alpha}^0$ to be an arbitrary random vector chosen over interval [-1, 1].

**Loop**: Set $j = 0$ and do

1. For $k = 1, ..., K$, compute
$$w_k^j = ((\bar{\alpha}_k^j)^2 + \varepsilon)^{-1/2}$$

2. Find $\bar{\alpha}^{j+1}$ that minimize $\sum_{m=1}^{M} |z_m - \bar{z}_m|^2 + \mu \|W^{1/2}\bar{\alpha}\|_2$ with
$$w_k = w_k^j$$

3. If $\|\bar{\alpha}^{j+1} - \bar{\alpha}^j\| / \|\bar{\alpha}^j\| \leq err$, exit.

4. Set $j = j + 1$.

**Result**: Output the estimated $\bar{\alpha}^j$.

Minimization of $\sum_{m=1}^{M} |z_m - \bar{z}_m|^2 + \mu \|W^{1/2}\bar{\alpha}\|_2$ is a nonlinear least squares problem. In the implementation, we use the TRESNEI code in [Morini & Porcelli, 2010] to find its solution. The TRESNEI code implements the method which has global or strong local convergence properties in [Macconi et al, 2009]. In the simulation experiments, we find that the probabilities converging to the local minima increases with the increase of the sparse levels.

**ACKNOWLEDGEMENT**

We are grateful to the anonymous reviewers for their thoughtful comments and suggestions, which greatly improve our paper.

This work was partially supported by the National Science Foundation of China (60971090, 61171166, 61101193) and the Intelligent Engineering Program of Jiangsu Province.

**REFERENCES**

Blumensath, T. [2010] "Compressed Sensing with Nonlinear Observations," *http://eprints.soton.ac.uk/164753/1/B_Nonlinear.pdf*

Candès, E. J. & Tao, T. [2006] "Near optimal signal recovery from random projections: Universal encoding strategies," *IEEE Trans. on Inform. Theory,* **52**, 5406-5425.

Candès, E. J., Wakin, M. B. & Boyd, S. P. [2008] "Enhancing sparsity by reweighted $l_1$ minimization," *J. Fourier Anal Appl*, **14**, 877-905.

Daubechies, I., DeVore, R., Fornasier, M. & Gunturk, C. S. [2010] "Iteratively re-weighted least squares minimization for sparse recovery," *Comm. Pure Appl. Math.*, **63**, 1-38.

Dedieu, H. & Ogorzałek, M. J. [1997] "Identifiability and Identification of Chaotic Systems Based on Adaptive Synchronization," *IEEE Trans. Circ. Syst.-I: Fund. Th. Appl.*, **44**, 976-988.

Donoho, D. L. [2006] "Compressed sensing," *IEEE Trans. on Inform. Theory,* **52**,1289-1306.

Floriane, A., Gérard, B., Gilles, M. & Lilianne D.V. [2008], "Identifiability of discrete-time nonlinear systems: The local state isomorphism approach," *Automatica*, **44**, 2884-2889

Gleick, J. [1987] *Chaos: Making a New Science*, (Vintage Publisher).

Itoh, M., Yang, T. & Chua, L. O. [2001] "Conditions for impulsive synchronization of chaotic and hyperchaotic systems," *Int J. Bifurcation and Chaos,* **11**, 551-560.

Liu, X. Z. [2001] "Impulsive stabilization and control of chaotic systems", *Non-Linear Analysis,* **47**, 1081-1092.

Macconi, M., Morini, B. & Porcelli, M. [2009], "Trust-region quadratic methods for nonlinear systems of mixed equalities and inequalities," *Applied Numerical Mathematics*, **59**, 859-876.

Madsen, K., Nielsen, H.B. & Tingleff, O. [2004] *Methods for Non-Linear Least Squares Problems*, Technical University of Denmark, Lecture notes, available at http://www.imm.dtu.dk/courses/02611/nllsq.pdf.

Morini, B. & Porcelli, M. [2010] "TRESNEI, A MATLAB trust-region solver for systems of nonlinear equalities and inequalities," *Comput. Optim. Appl.*, DOI: 10.1007/s10589-010-9327-5.

Nguyen, L.T., Dinh, V.P., Zahir, M. H., Huu, T. H., Morgan, V. L. & Core. J.C. [2008] "Compressed Sensing using Chaos Filters," *Telecommunication Networks and Applications Conference (ATNAC 2008)*, 219-223.

Schmidt, M., Fung, G. & Rosaless, R. [2007] "Fast optimization methods for $l_1$ regularization: A comparative study and two new approaches," Lecture Notes In Artificial Intelligence, **4701**, 286-297.

Schmidt, M., Fung, G. & Rosaless, R. [2009] "Optimization Methods for L1-Regularization," UBC Technical Report TR-2009-19.

Tropp, J. A. [2006] "Random filtering for compressive sampling," *40th Annual Conf. on Information Science and Systems*, 216-217.

Tropp, J. A., Wakin, M. B., Duarte, M. F, Baron, D. & Baraniuk, R. G. [2006] "Random filters for compressive sampling and reconstruction," *Proc. Int. Conf. Acoust. Speech Signal Proc.*, **3**, 872-875.

Tropp, J. A., Laska, J. N., Duarte, M. F., Romberg, J. K. & Baraniuk, R. G. [2010] "Beyond Nyquist: Efficient Sampling of Sparse Bandlimited Signals," *IEEE Trans. Inform. Theory*, **56**, 520-544.

Tseng, P. & Yun, S. [2009] "Block-Coordinate Gradient Descent Method for Linearly Constrained Nonsmooth Separable Optimization," *J. Optim. Theory Appl.*, **140**, 513-535.

Yang, T. & Chua, L. O. [1997] "Impulsive stabilization for control and synchronization of chaotic systems: Theory and application to secure communication," *IEEE Trans. Circ. Syst.-I: Fund. Th. Appl.*, **44**, 976-988.

Yu, L., Barbot, J. P., Zheng, G. and Sun, H. [2010] "Compressive Sensing With Chaotic Sequence," *IEEE Signal Proc. Letters*, **17**, 731-734.